\documentclass{article}

\usepackage[preprint]{neurips_2019}

\usepackage[utf8]{inputenc} 
\usepackage[T1]{fontenc}    
\usepackage{hyperref}       
\usepackage{url}            
\usepackage{booktabs}       
\usepackage{amsfonts}       
\usepackage{nicefrac}       
\usepackage{microtype}      

\title{Revised Progressive-Hedging-Algorithm Based Two-layer Solution Scheme \\for Bayesian Reinforcement Learning}

\author{%
  Xin Huang\\
  The Chinese University of Hong Kong\\
  \texttt{huangxin@se.cuhk.edu.hk}\\
  \And
  Duan Li\thanks{Corresponding author.}\\
  City University of Hong Kong\\
  \texttt{dli226@cityu.edu.hk}\\
  \And
  Daniel Z.~Long\\
  The Chinese University of Hong Kong\\
  \texttt{zylong@se.cuhk.edu.hk}\\
}

\usepackage{natbib}
\usepackage{amsfonts}
\usepackage{amsmath}
\usepackage{graphicx}
\usepackage{subfigure}
\usepackage[ruled,noend]{algorithm2e}
\usepackage{booktabs}
\usepackage{multirow}
\usepackage{stfloats}
\usepackage{color,soul} 
\usepackage{caption}

\begin{document}

\bibliographystyle{apalike}
\setcitestyle{round,colon,aysep={,},yysep={;}}
\renewcommand\refname{}

\maketitle

\begin{abstract}
Stochastic control with both inherent random system noise and lack of knowledge on system parameters constitutes the core and fundamental topic in reinforcement learning (RL), especially under non-episodic situations where online learning is much more demanding. This challenge has been notably addressed in Bayesian RL recently where some approximation techniques have been developed to find suboptimal policies. While existing approaches mainly focus on approximating the value function, or on involving Thompson sampling, we propose a novel two-layer solution scheme in this paper to approximate the optimal policy directly, by combining the time-decomposition based dynamic programming (DP) at the lower layer and the scenario-decomposition based revised progressive hedging algorithm (PHA) at the upper layer, for a type of Bayesian RL problem. The key feature of our approach is to separate reducible system uncertainty from irreducible one at two different layers, thus decomposing and conquering.  We demonstrate our solution framework more especially via the linear-quadratic-Gaussian problem with unknown gain, which, although seemingly simple, has been  a notorious subject over more than half century in dual control.
\end{abstract}

\section{Introduction}

Traditional reinforcement learning (RL) algorithms have been mainly developed under episodic settings, where learning can be done \emph{offline} by repeated simulations of an unchanged system \citep{sutton2018reinforcement,bertsekas2019reinforcement}. However, RL under a non-episodic problem setting with uncertainties resulted from both inherent systems random disturbance and lack of knowledge on system parameters still remains challenging, since such a problem requires learning \emph{online} and the exploitation-exploration tradeoff under such a setting becomes much more crucial and prominent. Bayesian RL has been thus emerged to tackle this essential issue in the machine learning community [Please see a comprehensive survey by \citep{ghavamzadeh2015bayesian}]. By augmenting the state space with probabilistic belief on those unknown systems parameters, the corresponding Bellman equation naturally optimizes the tradeoff between the exploitation by utilizing the current information from \emph{hyperstate} and exploration by foreseeing the outcomes of applied future actions using the Bayesian law, which leads to \emph{active learning}. As a consequence, the resulting current action from DP takes into account not only the current state and belief but also the future observations as future actions' outcomes by \emph{conditional planning} based on the posterior belief \citep[see][]{poupart2006analytic}.

Actually, a similar subject termed \emph{dual control} was invented long ago in \citet{feldbaum1960-1961dual}. Compared with classical adaptive controllers, the dual controller, as a function of the hyperstate, possesses dual features of both \emph{caution} to control the process (i.e., exploitation) and \emph{probing} to learn (identify) the system (i.e., exploration) \citep[see][]{bar1981stochastic}. No matter whether in Bayesian RL or dual control, however, to find an optimal policy is always analytically intractable, simply because of the very nonlinearity of inference, even under the Gaussian assumption \citep{aoki1967optimization}. It has been a long standing research challenge for over a half century in dual control, even for the ``toy'' example of discrete-time linear-quadratic-Gaussian (LQG) control problem with unknown gain. The subject becomes hot again recently in RL as evidenced by increasing attention in the community. For instance, \citet{dallaire2009bayesian} provide an algorithm within Bayesian RL framework in order to learn an LQG system without assuming any parametric form for the transition, observation and reward function. As they mention, however, it is hard to evaluate the control performance of their algorithm and compare with others. \citet{ouyang2017learning} address LQG case with unknown system parameters from Thompson sampling \citep{thompson1933likelihood} perspective. Such an approach of making decision based on one sample from current probabilistic model dose not take into account the learning effect of future actions. Researchers in RL, for example \citet{klenske2016dual} among others, became aware of some early work in dual control \citep[such as][]{tse1973actively}. It turns out that such algorithms are mostly classified as approximation on the inference for Bayesian RL. Our work in this paper is different from the existing literature in both RL and dual control. We propose for a type of Bayesian RL problem a novel solution framework based on techniques from progressive hedging algorithm (PHA) by \citet{rockafellar1991scenarios}. More precisely, we consider in this paper a discrete-time LQG problem with fixed but unknown gain in system dynamics, as a starting point in developing novel RL solution algorithms for this general class of problem. We would also like to emphasize that a discrete-time LQG problem with fixed but unknown gain in system dynamics is closely related to a well-studied dynamic portfolio selection problem, see for example \citet{li2000optimal}. We adopt a two-layer solution scheme, where at the lower layer, we solve a family of reduced stochastic control problem involving only irreducible uncertainty (i.e., random system noise) while fixing the reducible uncertainty (i.e., unknown parameter) at one possible value. For LQG case, the lower-layer problem can be analytically solved by dynamic programming (DP). At the upper layer, we apply a revised version of PHA in \citet{rockafellar1991scenarios} to iteratively aggregate the family of optimal policies from the lower layer to an implementable policy with respect to the posterior probabilities updated from nominal trajectories, which is motivated by \citet{li2008optimal}. We finally show in Section \ref{4} by experiments that our new approach leads to a better approximate policy with lower average cost than the outcome generated by DUL (\citep{deshpande1973adaptive}) which is still a leading \emph{passive learning} method in dual control up to date, and the other three methods using ideas borrowed from greedy method, $\epsilon$-greedy and Thompson sampling.

The rest of the paper is arranged as follows. We present in Section 2 the mathematical formulation of RL problem concerned. We then propose our two-layer scheme in Section 3 with an illustrative example, followed by the experimental results on performance comparing with other algorithms in Section 4. We conclude our paper in Section 5.
\section{Problem formulation}
\subsection{Bayesian reinforcement learning problems under consideration}

We consider a class of Bayesian RL problem with the state dynamics
$
x_{t+1} =f_t(x_t,u_t,\xi_t|\theta)
$
for $t=0,1,\cdots,T-1$ and $x_0$ given,
where $T$ is the finite time horizon, $x_t\in\mathbb{R}^n$ is the state which is assumed to be perfectly observed at time $t$, and $u_t\in\mathbb{R}^m$ is the control. The state transition function $f_t$ can be nonlinear in general, and is determined by a \emph{fixed} parameter $\theta\in\mathbb{R}^d$ which is, however, \emph{unknown} to the agent, who only has some prior belief on its distribution $p_0(\theta)$. Noise $\xi_t\in\mathbb{R}^n$ is the i.i.d. Gaussian disturbance with mean $0$ and covariance $\Sigma_\xi$, and is independent of $\theta$.

Given the above system, the goal of the agent is to minimize the following expected total cost,
\[
\mathcal{L}\big(\boldsymbol u|x_0,p_0(\theta)\big)=\mathbb{E}_{\theta,\boldsymbol{\xi}}\left[g_T(x_T)+\sum\limits_{t=0}^{T-1}g_t(x_t, u_t,\xi_t)\bigg|x_0,p_0(\theta)\right]
\]
over all admissible policies $\boldsymbol u=(u_0,u_1,\cdots,u_{T-1})'\in\mathcal{U}_0\times\mathcal{U}_1\times\cdots\times\mathcal{U}_{T-1}$, where $g_t$'s and $g_T$ are cost functions. We also assume that when fixing $\theta$ at $\theta_i$, $\mathcal{L}(\cdot|x_0, \delta(\theta=\theta_i))$ is convex w.r.t. $\boldsymbol u$, as required by PHA. In the following we denote $\delta(\theta=\theta_i)$ by $\delta_i$.

Note that the uncertainty from $\theta$ is due to the lack of knowledge of the agent, and such an uncertainty is reducible by learning, whereas the randomness incurred by system disturbance $\xi_t$'s is not. Let us define the information set as
$I^t$ = $\{x_0, x_1, \ldots, x_t, u_0, u_1, \ldots, u_{t-1}\}$,~$t$ = 1, $\cdots$, $T-1$,
with $I^0 = \{p_0(\theta),x_0\}$, and denote the posterior distribution of $\theta$ at time $t$ by $p_{t}(\theta|I^t)$. Then the Bayesian law leads to 
\begin{eqnarray}
p_{t}(\theta|I^t)=p_{t}(\theta|x_t,u_{t-1},I^{t-1})\propto
\psi(x_t|\theta,x_{t-1},u_{t-1})\times p_{t-1}(\theta|I^{t-1}), 
\label{belief_proportion}
\end{eqnarray}
for $t=1,\cdots,T-1$ with $p_{0}(\theta|I^0)=p_{0}(\theta)$. For simplicity, we denote $p_{t}(\theta|I^t)$ by $p_{t}(\theta)$, but keep in mind that the calculation on belief of $\theta$ at time $t$ is $I^t$ dependent. Moreover, the conditional density of $x_t$ is given by
$
\psi(x_t|\theta,x_{t-1},u_{t-1}) = \phi(f_{t-1}^{-1}(x_t|\theta,x_{t-1},u_{t-1});0, \Sigma_\xi)|\text{det}(\mathcal{J}_{f_{t-1}^{-1}}(x_t))|
$
with $\phi(z; \mu, \Sigma)$ being the normal density at $z$ with mean $\mu$ and variance $\Sigma$, and $\mathcal{J}$ being the Jacobian matrix of $f_{t-1}^{-1}(x_t|\theta,x_{t-1},u_{t-1})$. We thus find the recursive relationship between $p_{t+1}(\theta)$ and $p_{t}(\theta)$. When implementing controls successively, the agent will update the belief on $\theta$ accordingly after observing the realized state, and, hopefully, the agent's knowledge of $\theta$ will \emph{degenerate} when the time horizon $T$ is large enough.
By \emph{augmenting} the original state space with belief of model parameters, the Bellman equation for the optimal value function $J_t$ of the augmented system is given by
\begin{eqnarray}
J_t\big(x_t,p_t(\theta)\big) = \min_{u_t}~\mathbb{E}_{\theta,\xi_t}\Big[g_t(x_t, u_t,\xi_t)+J_{t+1}\big(x_{t+1},p_{t+1}(\theta)\big)\big|x_t,p_t(\theta)\Big],
\label{general bellman}
\end{eqnarray}
for $t=0,\cdots,T-1$ with terminal condition $J_T(x_T,p_T(\theta))=g_T(x_T)$. Theoretically, we could handle Bayesian RL problems by DP, through augmenting the state space and keeping track of belief updating, thus achieving optimal balance between exploitation and exploration. The resulting optimal policy, if we are able to derive it, includes an essential feature of \emph{active} learning, in the sense that taking into account in (\ref{general bellman}) the effect of future actions and beliefs via conditional planning \emph{before} we actually observe the future states. Unfortunately, due to the high nonlinearity of (\ref{belief_proportion}), solving the Bellman equation (\ref{general bellman}) in general is impossible, at least intractable.

\subsection{LQG problem with unknown gain}\label{2.2}

As a specific case of the general form described in the above subsection, we consider in this subsection the following discrete-time LQG control problem with an unknown gain,
 \begin{eqnarray}
 \small
 (\mathcal{P})&~&\min\limits_{\boldsymbol u}~\mathbb{E}_{\theta,\boldsymbol{\xi}}\left[{\frac{1}{2}}x_T'Qx_T+\sum_{t=0}^{T-1} \Big(\frac{1}{2}x_t'Qx_t + \frac{1}{2}u_t'Ru_t \Big)\Big|x_0,p_0(\theta)\right]\nonumber\\
 &~&{\rm s.t.}~~x_{t+1} = Ax_t+B(\theta)u_t+\xi_t,~t=0,1,\cdots,T-1,\nonumber
  \end{eqnarray}
where $A$ is given, and the gain matrix $B$ is an {\it unknown} constant matrix determined by an unknown parameter $\theta$. The LQG problem under such an assumption is not trivial, as \citet{aastrom1986dual} pointed out more than three decades ago, and still remains as an open challenge, evidenced by the recent work of \citet{klenske2016dual} in RL community, in which the authors developed an approximate dual method for LQG problems with unknown parameters. Moreover, this type of problem is also of great practical interests. For instance, in the dynamic portfolio selection problem in \citet{li2000optimal}, the wealth dynamics is governed by a multiple-input and single-output linear system with the wealth level being the state and dollar amounts to invest on multiple risky assets being the control. The scalar $A$ in such a case is the deterministic risk-free rate, while the vector $B$ is the excess return vector of \emph{risky} assets. As an \emph{extension} of \citet{li2000optimal}, $B$ can be decomposed into an unknown but fixed mean (depending on whether the market is bull,  bear or others) and an i.i.d. Gaussian noise. What's more, due to the dynamic nature of the market, making a financial investment decision \emph{cannot} be an episodic experiment.

We assume in this paper that $\theta$ takes one of $N$ possible values, $\theta_1, \theta_2, \ldots, \theta_N$, with prior belief $p_0(\theta)$ = $(p_{01}, p_{02}, \ldots, p_{0N})'$ where $p_{0i}=\mathbb{P}(\theta= \theta_i \mid I^0)$, $i=1,\cdots,N$.
For simplicity we set
$
B(\theta_i) = B_i
$
for all $i$. Let the posterior probability of $\theta_i$ at time $t+1$ be
$
p_{(t+1)i} = \mathbb{P}(\theta = \theta_i \mid I^{t+1}).
$
Then, by the Bayesian law, we have, for $t=0,\cdots,T-1$,
\begin{eqnarray}
p_{(t+1)i} = \frac{\phi(x_{t+1}; Ax_{t} + B_iu_{t}, \Sigma_\xi) p_{ti}}{\sum_{j=1}^N [\phi(x_{t+1}; Ax_{t} + B_ju_{t}, \Sigma_\xi) p_{tj}]}
\label{LQG belief}
\end{eqnarray}
which is very nonlinear with respect to the realized $x_{t+1}$. Furthermore, the Bellman equation becomes
\begin{eqnarray}
J_t\big(x_t,p_t(\theta)\big)=\min_{u_t}~\mathbb{E}_{\theta,\boldsymbol{\xi}}\left[\displaystyle\frac{1}{2}x_t'Qx_t+ \frac{1}{2}u_t'Ru_t+J_{t+1}\big(x_{t+1},p_{t+1}(\theta)\big)\Big|x_t,p_t(\theta)\right]
\label{LQG bellman}
\end{eqnarray}
with terminal condition $J_T\big(x_T,p_T(\theta)\big)=\frac{1}{2}x_T'Qx_T$. It is easy to see that the value function is no longer quadratic, and DP would fail to deliver an analytical solution.
\section{A two-layer solution scheme}
We propose now our two-layer solution scheme for the class of Bayesian RL problem presented in the previous section. The basic idea is to separate reducible and irreducible uncertainties into two different layers, first decompose and then conquer. For simplicity and clearness of our presentation, we focus our investigation to the LQG problem with unknown gain described in Subsection \ref{2.2}.
\subsection{Classical PHA}
While DP is a time-decomposition based algorithm, the progressive hedging algorithm (PHA) by \citet{rockafellar1991scenarios} is a scenario-decomposition based scheme. PHA has been developed for solving multistage stochastic decision-making problems with {\it finite} scenarios. We now demonstrate how we can apply PHA to solve a {\it naive version} of problem $(\mathcal{P})$ in which the prior knowledge of parameter $\theta$ is not updated from learning. Note that the possible value of $\theta$ takes from a finite set, with a finite scenario index set $S=\{1,2,\cdots,N\}$. When we set $\theta=\theta_i$ hence $B=B_i,~i\in S$,  the related scenario subproblem is
 \begin{eqnarray}
 \small
 (\mathcal{P}_i)&~&\min\limits_{\boldsymbol u}~\mathcal{L}(\boldsymbol u|x_0,\delta_i):=\mathbb{E}_{\boldsymbol{\xi}}\Big[\displaystyle{\frac{1}{2}}x_T'Qx_T+\sum\limits_{t=0}^{T-1} \Big(\frac{1}{2}x_t'Qx_t+ \frac{1}{2}u_t'Ru_t \Big)\Big|x_0\Big] \nonumber\\
 &~&{\rm s.t.}~~x_{t+1} = Ax_t+B_i u_t+\xi_t,~t=0,1,\cdots,T-1,\nonumber
  \end{eqnarray}
which returns to a classical LQG problem with \emph{known} gain and irreducible uncertainty associated with $\{\xi_t\}$. Scenario subproblem $(\mathcal{P}_i)$ can be solved analytically by DP with the following optimal feedback policy at time $t$ \citep[refer to][for example]{kirk1970optimal}
\begin{eqnarray}
u^*_{ti}(x_t)=-K_{ti}x_{t},
\label{LQG_classical_solution}
\end{eqnarray}
where
$K_{ti}=(R+B_i'P_{(t+1)i}B_i)^{-1}B_i'P_{(t+1)i}A$
and
$P_{ti} = Q + K_{ti}'RK_{ti} + (A-B_iK_{ti})'P_{(t+1)i}(A-B_iK_{ti})$. If we do not update the prior knowledge of parameter $\theta$ using future observations, we have a {\it naive version} of problem $(\mathcal{P})$,
\begin{eqnarray}
 (\mathcal{P}_{Naive})&~&\min\limits_{\boldsymbol u}~\textstyle\sum_{i\in S} p_{0i}\mathcal{L}(\boldsymbol u|x_0,\delta_i)~~~{\rm s.t.}~x_{t+1} = Ax_t+B(\theta) u_t+\xi_t,~t=0,1,\cdots,T-1.\nonumber
\end{eqnarray}
To solve $(\mathcal{P}_{Naive})$, PHA, as a \emph{scenario-decomposition} method, first decomposes it into $N$ scenario subproblems $(\mathcal{P}_{i})$, $i\in S$, and generates the scenario-based feedback policy for each $i$ using (\ref{LQG_classical_solution}),
\begin{eqnarray}
\boldsymbol{u}^{[0]}_i(\cdot)=\big(u_{0i}^{*}(x_0),u_{1i}^{*}(x_1),\cdots,u_{(T-1)i}^{*}(x_{N-1})\big)',
\label{scenario_sub_sol}
\end{eqnarray}
which does not satisfy the \emph{non-anticipativity}. To proceed, PHA projects all $\boldsymbol{u}^{[0]}_i$'s into a non-anticipative space to get an \emph{implementable} feedback policy
$
\boldsymbol{\hat u}^{[0]} = \sum_{i\in S} p_{0i}\boldsymbol{u}^{[0]}_i.
$
Compared with a scenario-specific policy (like $\boldsymbol{u}^{[0]}_i$), an implementable one (like $\boldsymbol{\hat u}^{[0]}$) is \emph{indifferent} to all scenarios. In order to generate the final \emph{optimal} implementable feedback policy, PHA next adopts the augmented Lagrangian method by adding some appropriate penalties to the original scenario subproblems. More specifically, PHA solves, in parallel for all $i$ $\in$ $S$, the resulting Lagrangian subproblems
\begin{eqnarray}
 (\mathcal{P}_i^{[\nu]})~~~\min\limits_{\boldsymbol u}~\mathcal{L}(\boldsymbol u|x_0,\delta_i)+\boldsymbol u'\boldsymbol{w}_i^{[\nu]}+\frac{1}{2}r\|\boldsymbol u-\boldsymbol{\hat u}^{[\nu]}\|^2,\nonumber
\end{eqnarray}
in each iteration $\nu=0,1,\cdots$, whose optimal solution is denoted by $\boldsymbol{u}^{[\nu + 1]}_i$, with the next implementable feedback policy used for $(\mathcal{P}_i^{[\nu+1]})$ being
$
\boldsymbol{\hat u}^{[\nu + 1]}=\sum_{i\in S} p_{0i}\boldsymbol{u}^{[\nu + 1]}_i.
$
Furthermore, the Lagrangian multiplier for scenario $i$ is updated via
$
\boldsymbol{w}_i^{[\nu+1]}=\boldsymbol{w}_i^{[\nu]}+r\big(\boldsymbol{u}^{[\nu+1]}_i-\boldsymbol{\hat u}^{[\nu+1]}\big)
$
with initial $\boldsymbol{w}_i^{[0]}$ being zero vector for all $i$; the penalty parameter $r>0$ is predetermined, and $\|\cdot\|$ denotes $2$-norm hereafter. The process repeats until the convergence occurs, which is guaranteed by a convexity of the scenario subproblem w.r.t. the control variable according to the PHA requirement, and is actually the optimal solution to the problem $(\mathcal{P}_{Naive})$.
\subsection{Revised PHA-based two-layer solution scheme}
In order to incorporate learning feature into our solution algorithm for $(\mathcal{P})$, we need to update the knowledge about uncertain parameter $\theta$. In our revised PHA-based two-layer solution scheme, every step is the same as in PHA for the naive version $(\mathcal{P}_{Naive})$ described in the previous subsection, except that, when forming the implementable feedback policy at iteration $\nu$, we need to take conditional expectations using the \emph{posterior} probabilities at each time $t$ = 0, 1, $\cdots$, $T-2$,
\begin{eqnarray}
{\hat u_{t+1}}^{[\nu]}\left(\cdot\right)=\textstyle\sum_{i\in S} p_{(t+1)i}u_{(t+1)i}^{[\nu]}\left(\cdot\right),~
\label{general_projection}
\end{eqnarray}
where $u_{(t+1)i}^{[\nu]}\left(\cdot\right)$ of $\boldsymbol u^{[\nu]}_i(\cdot)$ comes from solving the $i$th Lagrangian subproblem $(\mathcal{P}_i^{[\nu-1]})$ ($\nu\geq 1$), with initial $\boldsymbol u^{[0]}_i$ given by (\ref{scenario_sub_sol}). As evidenced from (\ref{LQG belief}), $p_{(t+1)i}$ depends on $I^{t+1}$. If we directly substitute (\ref{LQG belief}) into (\ref{general_projection}), ${\hat u_{t+1}}^{[\nu]}(\cdot)$ becomes \emph{nonlinear} in state, which in turn leads to the intractability when dealing with $(\mathcal{P}_i^{[\nu]})$ in the next iteration. We bypass this difficulty by setting $p_{(t+1)i}$ at its \emph{nominal} value,
\begin{eqnarray}
\bar p_{(t+1)i}^{[\nu]} = \frac{\phi(\bar x_{t+1}^{[\nu]}; \bar\mu_{ti}^{[\nu]}, \Sigma_\xi) \bar p_{ti}^{[\nu]}}{\sum_{j\in S} [\phi(\bar x_{t+1}^{[\nu]}; \bar\mu_{tj}^{[\nu]}, \Sigma_\xi) \bar p_{tj}^{[\nu]}]}
\label{LQG nominal_belief}
\end{eqnarray}
where $\bar\mu_{ti}^{[\nu]}=A\bar x_{t}^{[\nu]} + B_i\hat u_{t}^{[\nu]}(\bar x_{t}^{[\nu]})$ and the \emph{nominal} state is determined sequentially by
\begin{eqnarray}
\bar x_{t+1}^{[\nu]}&=&\mathbb{E}_{\theta,\xi_t}\big[A\bar x_{t}^{[\nu]}+B(\theta)\hat u_t^{[\nu]}(\bar x_{t}^{[\nu]})+\xi_t~\big|~\bar x_{t}^{[\nu]},\bar p_{t}^{[\nu]}(\theta)\big]\nonumber\\
&=&A\bar x_{t}^{[\nu]}+\big(\textstyle\sum_{i\in S}\bar p_{ti}^{[\nu]}B_i\big)\hat u_t^{[\nu]}(\bar x_{t}^{[\nu]})
\label{nominal_state}
\end{eqnarray}
for $t=0,1,\cdots,T-1$ with nominal initial state $\bar x_{0}^{[\nu]}=x_0$ and nominal prior distribution $\bar p_{0}^{[\nu]}(\theta)=p_{0}(\theta)$ held for every $\nu$. Then the implementable feedback policy obtained at $t+1$ becomes \emph{linear} with respect to the state compared with (\ref{general_projection}),
\begin{eqnarray}
{\hat u_{t+1}}^{[\nu]}\left(\cdot\right)=\textstyle\sum_{i\in S} \bar p_{(t+1)i}^{[\nu]}u_{(t+1)i}^{[\nu]}\left(\cdot\right).~
\label{numerical_projection}
\end{eqnarray}
This relaxation to a linear policy enables us to proceed the iteration until converging to a final approximate feedback policy of our two-layer (TL) method,
\begin{eqnarray}
u_t^{TL}(x_t)=\hat u_{t}^{[\nu]}(x_t)=-\hat K_{t}^{[\nu]}x_t,~\text{as}~\nu\rightarrow\infty,~\text{for all }t.
\label{TL_policy}
\end{eqnarray}
In practice, the algorithm will stop when the predetermined convergence tolerance level ($tol$) is satisfied, namely, $err < tol$, where the error is defined by
\begin{eqnarray}
err:=\textstyle\sqrt{\|\boldsymbol{\hat u}^{[\nu+1]}-\boldsymbol{\hat u}^{[\nu]}\|^2+\frac{1}{r^2}\sum_{i\in S}\|\boldsymbol{w}_i^{[\nu+1]}-\boldsymbol{w}_i^{[\nu]}\|^2}.
\label{err}
\end{eqnarray}
We summarize the complete algorithm as the pseudocode in Algorithm \ref{TL_algo}.

\begin{algorithm}[htbp]
\small
\caption{Revised PHA-based Two-layer Algorithm for LQG with An Unknown Gain
\label{TL_algo}}
\KwIn{System known parameters and functions, the initial state, the prior belief of $\theta$ with the scenario index set $S=\{1,\cdots,N\}$, the algorithm tolerance level $tol$, and the penalty $r$}
\KwOut{Approximate feedback policy (suboptimal)}
\emph{1. Lower layer: deal with scenario subproblems.}\\
solve $(\mathcal{P}_i)$ to get $\boldsymbol{u}^{[0]}_i$ for all $i\in S$;\\
\emph{2. Upper layer: active learning along a nominal trajectory.}\\
\emph{2.1} Initialize the implementable feedback policy $\boldsymbol{\hat u}^{[0]}$ and the Lagrangian multiplier $\boldsymbol{w}_i^{[0]}$ in order to form the initial Lagrangian subproblems $(\mathcal{P}_i^{[0]})$:\\
\While{$\nu=0$}{
    set the nominal initial state $\bar x_{0}^{[0]}=x_0$ and the nominal prior distribution $\bar p_{0}^{[0]}(\theta)=p_{0}(\theta)$;\\
    obtain $\boldsymbol{\hat u}^{[0]}=\{{\hat u}^{[0]}_0,{\hat u}^{[0]}_1,\cdots,{\hat u}^{[0]}_{T-1}\}$ through \emph{forward} calculation:\\
    \For{$t:0\rightarrow (T-1)$}{
        compute the implementable feedback policy at $t$: ${\hat u_{t}}^{[0]}\left(\cdot\right)=\sum\limits_{i\in S} \bar p_{ti}^{[0]}\times u_{ti}^{[0]}\left(\cdot\right)$;\\
        compute the nominal state of next time stage $\small\bar x_{t+1}^{[0]}$ via (\ref{nominal_state}) for $\nu=0$;\\
        update the nominal posterior distribution $\bar p_{t+1}^{[0]}(\theta)$:\\
        \For{$i\in S$}{
            calculate scenario-specific control $\bar\mu_{ti}^{[0]}=A\bar x_{t}^{[0]} + B_i\hat u_{t}^{[0]}(\bar x_{t}^{[0]})$;\\
            update nominal posterior probability $\small\bar p_{(t+1)i}^{[0]}$ via (\ref{LQG nominal_belief}) for $\nu=0$;\\
        }
    }
}
set $\boldsymbol{w}_i^{[0]}=\boldsymbol{0}$ for all $i\in S$;\\
\emph{2.2} Iteration starts:\\
\For{$\nu=0,1,\cdots$}{
    form the Lagrangian subproblems and solve in parallel: $(\mathcal{P}_i^{[\nu]})$ to obtain $\boldsymbol{u}^{[\nu+1]}_i$ for all $i\in S$;\\
    conduct the \textbf{while-do} in 2.1 above but for $\nu+1$ in order to get $\boldsymbol{\hat u}^{[\nu+1]}$;\\
    update the Lagrangian multiplier: $\boldsymbol{w}_i^{[\nu+1]}=\boldsymbol{w}_i^{[\nu]}+r\big(\boldsymbol{u}^{[\nu+1]}_i-\boldsymbol{\hat u}^{[\nu+1]}\big)$ for all $i\in S$;\\
    check the algorithm stopping criterion by calculating $err$ via (\ref{err}):\\
    \eIf{$err < tol$}{
        \textbf{break};
    }{
    $\nu \leftarrow \nu+1$;
    }
}
\end{algorithm}

Confining ourselves on the nominal trajectory, on the one hand, we are able to \emph{forwardly} calculate an implementable policy and the nominal posterior distribution along the time horizon in each iteration, on the other hand, however, since Bellman equation in (\ref{LQG bellman}) considers the entire (continuous) state space, the converged nominal-based policy is only suboptimal. Nevertheless, as we will demonstrate in Section \ref{4}, our newly-derived approximation performs better in an average sense than the prevalent passive learning method and others borrowed from traditional RL algorithms. Before that, let us take an example to further explain our algorithm.

\subsection{Illustration of the algorithm}\label{3.3}

In this subsection, we use a scalar system to illustrate Algorithm 1 in details. More specifically, the system dynamics becomes $x_{t+1} = ax_t+b(\theta)u_t+\xi_t$ for $t=0,1,\cdots,T-1$, where we denote $b(\theta_i)=b_i~,\forall i\in S=\{1,\cdots,N\}$ and the i.i.d. system random disturbance $\xi_t$ follows $\mathcal{N}(0,\sigma^2)$, together with other usual assumptions for $(\mathcal{P})$.

Solving the scenario subproblem $\min_{\boldsymbol u}~\mathcal{L}(\boldsymbol u|x_0,\delta_i)$ by DP for each $i$ at the lower layer, we obtain the scenario-specific feedback policy
$
u^{[0]}_{ti}(x_t)=-K_{ti}^{[0]}x_{t}
$
for all $t$, with backward recursions $K_{ti}^{[0]}=(ab_i\Gamma_{(t+1)i}^{[0]})/(R+b_i^2\Gamma_{(t+1)i}^{[0]})$, $\Gamma_{ti}^{[0]} = Q + R(K_{ti}^{[0]})^2 + (a-b_iK_{ti}^{[0]})^2\Gamma_{(t+1)i}^{[0]}$, and the boundary condition $\Gamma_{Ti}^{[0]} \equiv Q$ for all $i$. The optimal cost-to-go function is then given by
$
J_{ti}^{[0]}(x_t) = \frac{1}{2}\Gamma_{ti}^{[0]}x_t^2 + \Lambda_{ti}^{[0]},
$
with
$
\Lambda_{ti}^{[0]} = \Lambda_{(t+1)i}^{[0]} + \frac{1}{2}\Gamma_{(t+1)i}^{[0]}\sigma^2
$
and the boundary condition $\Lambda_{Ti}^{[0]}\equiv 0$. Then at the upper layer we first \emph{forwardly} aggregate solutions from the lower layer and simultaneously update the posterior distribution along the nominal state trajectory in order to form the initial Lagrangian subproblems $(\mathcal{P}_i^{[0]})$ for later iteration. More precisely, we initialize the nominal initial state $\bar x_{0}^{[0]}=x_0$ and the nominal prior distribution $\bar p_{0}^{[0]}(\theta)=p_{0}(\theta)$, and then calculate the implementable control at $t=0$:
$
\hat u_{0}^{[0]}(\bar x_{0}^{[0]})=\textstyle\sum_{i\in S} \bar p_{0i}^{[0]}u^{[0]}_{0i}(\bar x_{0}^{[0]})=-\hat K_{0}^{[0]}\bar x_{0}^{[0]}
$
where
$
\hat K_{0}^{[0]}=\textstyle\sum_{i\in S}\bar p_{0i}^{[0]}K_{0i}^{[0]}.
$
The next nominal state is obtained via (\ref{nominal_state}):
$
\bar x_{1}^{[0]}=\big[a-(\sum_{i\in S} \bar p_{0i}^{[0]}b_i)\hat K_{0}^{[0]}\big]\bar x_{0}^{[0]},
$
with the nominal posterior probabilities $\bar p_{1i}^{[0]}$'s updated \emph{numerically} through (\ref{LQG nominal_belief}) for various $i$, which in turn yields $\hat u_{1}^{[0]}(x_1)=-\hat K_{1}^{[0]}x_1$ with $\hat K_{1}^{[0]}=\sum_{i\in S} \bar p_{1i}^{[0]}K_{1i}^{[0]}$, which is still linear in state. We then apply the control $\hat u_{1}^{[0]}(\bar x_{1}^{[0]})$ to get the nominal state and posterior distribution at $t=2$. We conduct the above procedure till the end of time horizon and finally obtain a feedback policy $\boldsymbol{\hat u}^{[0]}$ for $\nu=0$. The initial Lagrangian multiplier is set to be zero, which is equivalent as a linear function of state, namely, $w^{[0]}_{ti}(x_t)=W^{[0]}_{ti}x_t$ with $W^{[0]}_{ti}\equiv0$ for all $t$ and $i$. We are now ready to define Lagrangian subproblems for $\nu=0$:
$
(\mathcal{Q}_i^{[0]}):~\min_{\boldsymbol u}~\mathcal{L}(\boldsymbol u|x_0,\delta_i)+\sum_{t=0}^{T-1}w^{[0]}_{ti}u_{t}+\frac{1}{2}r\sum_{t=0}^{T-1}(u_{t}-\hat u_{t}^{[0]})^2.\nonumber
$
Note that the above multistage optimization problem (subject to the linear state dynamics) can be \emph{analytically} solved by DP and the optimal solution is \emph{linear} in state. It is easy to prove, by mathematical induction, that starting with linear forms of $\hat u_{t}^{[0]}(\cdot)$ and $w^{[0]}_{ti}(\cdot)$, all $(\mathcal{Q}_i^{[\nu]})$'s, $\nu\geq0$, keep the same quadratic forms, with optimal cost-to-go function at $t$ satisfying the new Bellman equation
\begin{eqnarray}
\small
J_{ti}^{[\nu+1]}(x_t)=\min\limits_{u_{t}}~\mathbb{E}_{\boldsymbol{\xi}}\left[\displaystyle{\frac{1}{2}}Qx_t^2 + \displaystyle{\frac{1}{2}}Ru_t^2 + w^{[\nu]}_{ti}u_{t} + \displaystyle\frac{1}{2}r(u_{t}-\hat u_t^{[\nu]})^2+ J_{(t+1)i}^{[\nu+1]}\Big(ax_t+b_iu_{t}+\xi_t\Big)~\bigg|~x_t~\right],\nonumber
\end{eqnarray}
with boundary condition $J_{Ti}^{[\nu+1]}(x_{T})=\frac{1}{2}x_T^2$. Then optimal solution is $
u_{ti}^{[\nu+1]}(x_{t}) = -K_{ti}^{[\nu+1]}x_{t},
$
where
$
K_{ti}^{[\nu+1]} = (ab_i\Gamma_{(t+1)i}^{[\nu+1]}+r\hat K_{t}^{[\nu]}+W^{[\nu]}_{ti})/(R+r+b_i^2\Gamma_{(t+1)i}^{[\nu+1]}).
$
Accordingly,
$
J_{ti}^{[\nu+1]}(x_{t})=\frac{1}{2}\Gamma_{ti}^{[\nu+1]}x_{t}^2+\Lambda_{ti}^{[\nu+1]},
$
where
$
\Gamma_{ti}^{[\nu+1]}=\displaystyle a^2\Gamma_{(t+1)i}^{[\nu+1]}+Q+r\big(\hat K_{t}^{[\nu]}\big)^2-\big(R+r+b_i^2 \Gamma_{(t+1)i}^{[\nu+1]}\big)\big(K_{ti}^{[\nu+1]}\big)^2
$
and
$
\Lambda_{ti}^{[\nu+1]} = \Lambda_{(t+1)i}^{[\nu+1]}+{\frac{1}{2}}\Gamma_{(t+1)i}^{[\nu+1]}\sigma^2,
$
with boundary conditions $\Gamma_{Ti}^{[\nu+1]}=Q$ and $\Lambda_{Ti}^{[\nu+1]}=0$. Finally, the Lagrangian multiplier is updated via
$
w^{[\nu+1]}_{ti}(x_t)= w^{[\nu]}_{ti}(x_t)+r\big[u^{[\nu+1]}_{ti}(x_t)-\hat u^{[\nu+1]}_{t}(x_t)\big]=W^{[\nu+1]}_{ti}x_t,
$
where
$
W^{[\nu+1]}_{ti}=W^{[\nu]}_{ti}+r(\hat K_{t}^{[\nu+1]}-K_{ti}^{[\nu+1]}).
$
The iteration terminates when the stopping criterion described in Algorithm \ref{TL_algo} is satisfied, and results in a \emph{linear} feedback policy as in (\ref{TL_policy}).
\section{Experimental results on performance}\label{4}
By assigning different values for the \emph{non-episodic} example discussed in Subsection \ref{3.3}, we now verify the efficiency of our proposed two-layer (TL) scheme, compared with other algorithms including DP, DUL (the prevalent passive learning approach of \citet{deshpande1973adaptive}), and three other methods leveraging ideas from traditional RL algorithms: the greedy method, $\epsilon$-greedy, and Thompson sampling. While DP, as the theoretical best, provides a benchmark for comparison, it is \emph{only} applicable when $T=2$, where analytical optimal policy can be obtained at $t=1$, and \emph{numerical method} has to be invoked at $t=0$, for example by MATLAB. As for DUL, it makes an ansatz that the expectation and the minimization operators in the original problem $(\mathcal{P})$ can be exchanged, i.e.,
$
\small
\min_{\boldsymbol u}~\mathbb{E}_{\theta}\left\{\mathbb{E}_{\boldsymbol\xi}\left[~\cdots~\big|p_0(\theta)\right]\right\}\approx\mathbb{E}_{\theta}\big\{\min_{\boldsymbol u}~\mathbb{E}_{\boldsymbol\xi}\left[~\cdots~\big|p_0(\theta)\right]\big\}.
$
The DUL algorithm is basically a rolling horizon approach. At the beginning, scenario subproblems are solved for the entire time horizon and the resulting DUL policy at $t=0$ is a weighted sum of the optimal solutions for scenario subproblems with weighting coefficients being the prior probabilities. At time $t$, posterior distribution of $\theta$ is calculated based on observed $I^t$, and scenario subproblems are solved for the truncated time horizon from $t$ to $T$, and the resulting DUL policy at time $t$ is a weighted sum of the new scenario-based optimal solutions with weights being the posterior probabilities at time $t$. The essence of DUL is actually to take conditional \emph{average}. Adopting similar idea of rolling horizon, we may also think out other three algorithms (rooted originally in classical RL problems) that are applicable to non-episodic cases. The first one is similar to the \emph{greedy} method (named GRE) by selecting the scenario-specific policy with largest posterior probability at time $t$. As a variation of GRE, the \emph{$\epsilon$-greedy} type strategy (termed $\epsilon$-GRE) perturbs the greedy policy a bit by a randomized policy of selecting the greedy policy with probability $(1-\epsilon)$ or a randomly chosen policy with probability $\epsilon$. The last algorithm follows the idea of \emph{Thompson sampling} (labelled TS here) that a policy at time $t$ is selected by randomly sampling a scenario-specific policy based on the posterior distribution.

For simplicity, the model is set with $a=Q=R=\sigma=x_0=1$, and $N=2$ meaning that $b$ takes two possible values. The penalty parameter $r$ and the tolerance level are chosen to be 1 and $10^{-5}$, respectively. Details of assignments on $b=\{b_1,b_2\}$ and $p_0(\theta) =\{p_{01},p_{02}\}$ for 12 experiments can be found in Table \ref{experiment_table}. We do these for both simplest two-period problem, where DP works, and longer time horizon problems with $T=3$ and $T=5$, where DP fails. As for long-horizon problems with $T$ = 3 or 5, we adopt \emph{rolling} TL in order to incorporate real posterior belief into consideration. More specifically, at each time $t$ we adopt \emph{current} TL policy $\hat u_t^{TL}(\cdot)$ for only \emph{one} period and treat $p_{t+1}(\theta)$ based on the real-observed $I^{t+1}$ as ``prior'' belief in the next step to solve the remaining problem from $t+1$ to $T$ to generate a \emph{new} $\hat u_{t+1}^{TL}(\cdot)$ (termed as TL$_{R}$ method).

For each experiment under a certain group of $T$, $b$, and $p_0(\theta)$, we compute the TL feedback gain $\hat K_t$ in (\ref{TL_policy}). Then we generate ten thousand simulations for each experiment out of twelve shared by all the seven algorithms. Every simulation is characterized by two parts $(\theta, \{\xi_t\}_t)$, where $\theta$ is sampled by $p_0(\theta)$ and each $\xi_t$ is sampled from the assumed i.i.d. Gaussian noise, in order to calculate and compare the total costs induced by different policies in the \emph{average} sense. Table \ref{experiment_table} summarizes these results. When $T=2$, DP always ranks the top (with one exception, which could be due to that MATLAB can only identify a local minimum for a possible non-convex value function at $t$ = 0) and TL approximates the true optimal policy pretty well as evidenced by its lower average total cost compared to others (except for its rolling variant and DP). For $T=3$ where DP no longer works, TL almost maintains superior over the rest (except for TL$_{R}$) even without utilizing any \emph{online} posterior information which other approaches rely on. As time goes by, reference to newly-updated belief becomes more and more necessary. Based on this recognition, our TL$_{R}$ essentially beats all the rest when $T$ goes beyond 2. We also observe some interesting findings that should be naturally expected. First, the longer horizon, the larger the total cost. Second, in experiment $(vi)$ where we are \emph{certain} about the system parameter with a one-point distribution for the prior belief, all the algorithms lead to the same (actually optimal) policy and yield the same cost, since the problem, in such a case, reduces to a pure stochastic decision problem with \emph{full} knowledge on parameters. Finally, we can see from experiment $(i)$ to $(iii)$ that the larger the variance of $b$, the worse the passive learning DUL and others perform. In other words, the inherent active learning feature in TL (TL$_{R}$) and DP becomes much more demanding when the uncertainty in $b$ is large.

   \begin{table*}[htbp]
     \centering
     \caption{Average total costs from seven algorithms under 10,000 simulations for each experiment out of twelve, with different assignments on $T$, $b$ (with $N=2$ for simplicity), and $p_0(\theta)=\{p_{01},p_{02}\}$; $\epsilon$ is set to be 10\%; all the results are rounding in four decimals; \textbf{bold} numbers represent the lowest total costs under corresponding experiments.}
     \resizebox{1\textwidth}{!}{
       \begin{tabular}{ccclccccccc}
       \toprule
       No. & $p_{01}$ & \multicolumn{2}{c}{$b$} & DP & TL & TL$_{R}$ & DUL & GRE & $\epsilon$-GRE & TS \\
       \midrule
       \midrule
       \multicolumn{11}{l}{\emph{For $T=2$}}\\
       \midrule
       $(i)$ & \multirow{3}[2]{*}{$\displaystyle\frac{1}{3}$} & $b_1=1$ &  $b_2=2$ & \textbf{1.8170} & 1.8172 & 1.8171 & 1.8204 & 1.8213 & 1.8236 & 1.8409 \\
       $(ii)$ &    & $b_1=1$ &  $b_2=5$ & \textbf{1.8199} & 1.8261 & 1.8203 & 1.9974 & 2.0491 & 2.1060 & 2.4745 \\
       $(iii)$ &    & $b_1=1$ &  $b_2=10$ & 1.8793 & 1.8833 & \textbf{1.8785} & 2.7875 & 3.7598 & 4.0188 & 5.7482 \\
       \midrule
       $(iv)$ & $1/2$ & \multirow{3}[2]{*}{$b_1=1$} &  \multirow{3}[2]{*}{$b_2=2$} & \textbf{1.9052} & 1.9060 & 1.9055 & 1.9095 & 1.9296 & 1.9314 & 1.9310 \\
       $(v)$ & $2/3$ &  &  & \textbf{1.9383} & 1.9395 & 1.9389 & 1.9427 & 1.9614 & 1.9611 & 1.9589 \\
       $(vi)$ & $1$  &  &  & \textbf{2.0276} & \textbf{2.0276} & \textbf{2.0276} & \textbf{2.0276} & \textbf{2.0276} & \textbf{2.0276} & \textbf{2.0276} \\
       \midrule
       \midrule
       \multicolumn{11}{l}{\emph{For $T=3$}}\\
       \midrule
       $(vii)$ & $1/3$ & \multirow{3}[2]{*}{$b_1=1$} &  \multirow{3}[2]{*}{$b_2=2$} & N/A & 2.5349 & \textbf{2.5333} & 2.5371 & 2.5517 & 2.5545 & 2.5837 \\
       $(viii)$ & $1/2$ &  &  & N/A & 2.6541 & \textbf{2.6511} & 2.6542 & 2.6949 & 2.6916 & 2.6932 \\
       $(ix)$ & $2/3$  &  &  & N/A & 2.7140 & \textbf{2.7106} & 2.7139 & 2.7384 & 2.7419 & 2.7506 \\
       \midrule
       \midrule
       \multicolumn{11}{l}{\emph{For $T=5$}}\\
       \midrule
       $(x)$ & $1/3$ & \multirow{3}[2]{*}{$b_1=1$} &  \multirow{3}[2]{*}{$b_2=2$} & N/A & 3.8848 & \textbf{3.8779} & 3.8804 & 3.9070 & 3.9134 & 3.9671 \\
       $(xi)$ & $1/2$ &  &  & N/A & 4.0923 & \textbf{4.0762} & 4.0789 & 4.1359 & 4.1364 & 4.1559 \\
       $(xii)$ & $2/3$  &  &  & N/A & 4.2734 & \textbf{4.2546} & 4.2558 & 4.3022 & 4.3039 & 4.3159 \\
       \bottomrule
       \end{tabular}%
     }
     \label{experiment_table}%
     \vspace{-11pt}
   \end{table*}%

\section{Summary}

We develop a novel solution approach to a type of Bayesian reinforcement learning (RL) problem under the non-episodic setting, especially the discrete-time linear-quadratic-Gaussian (LQG) problem with fixed but unknown gain as one concrete example, to which the classical dynamic programming (DP) fails. While the existing algorithms in the RL literature focus mainly on approximating the value function when it comes to the single-episodic setting \citep[for example,][]{klenske2016dual}, or invoking Thompson sampling method in episodes for LGQ with uncertain parameters \citep[such as][]{ouyang2017learning}, our new solution approximates the optimal policy directly, thus bypassing the stage of approximating the value function. Most importantly, our scheme separates the non-episodic problem into two different layers according to different types of uncertainties, and combines the time-decomposition based method DP at the lower layer and the revised scenario-decomposition based approach progressive hedging algorithm (PHA) at the upper layer, to strike a balance between exploitation and exploration. By separating the reducible uncertainty from the irreducible one, we may take advantage of DP to generate an analytical solution for scenario-specific subproblems with reducible uncertainty fixed at a certain scenario. The revised PHA at the upper level, on the other hand, aggregates the solutions from all scenario subproblems to generate an implementable one, which finally converges to a suboptimal policy to approximate the optimal one of the primal Bayesian RL problem, as shown in our experiments. One future research topics are to investigate deeper the convergence property of our revised-PHA based two-layer solution algorithm, and to study how to generate nominal trajectory or even multiple ones in order to simulate more learning environment in advance. Furthermore, while the current version of PHA requires convexity for each scenario problem to guarantee the convergence, Prof. Rockafellar has recently been considering to relax PHA to nonconvex cases \citep{rockafellar2018progressive}. We will also utilize his new results to extend our current work.

\subsubsection*{References}
\vspace{-11pt}
\bibliography{NIPS2019_PHA}

\end{document}